\documentclass[prb,twocolumn,preprintnumbers,amsmath,amssymb]{revtex4}

\usepackage{epsfig,amsfonts}

\def\be{\begin{equation}}
\def\ee{\end{equation}}

\def\bea{\begin{eqnarray}}
\def\eea{\end{eqnarray}}

\def\e{\epsilon}

\def\a{\alpha}
\def\b{\beta}

\def\s{\sigma}

\def\ben{\begin{enumerate}}
\def\een{\end{enumerate}}

\newcommand{\bra}[1]{\left< #1 \right|}
\newcommand{\ket}[1]{\left| #1 \right>}
\newcommand{\vm}[1]{\left< #1 \right>}

\begin{document}

\title{Exact mesoscopic correlation functions of the pairing model}
\author{Alexandre Faribault${}^1$, Pasquale Calabrese${}^{2}$ and  
Jean-S\'ebastien Caux${}^{1}$}
\affiliation{$^1$Institute for Theoretical Physics, Universiteit van  
Amsterdam,
1018 XE Amsterdam, The Netherlands}
\affiliation{$^{2}$Dipartimento di Fisica dell'Universit\`a di Pisa and  
INFN,
56127 Pisa, Italy}

\date{\today}

\begin{abstract}
We study the static correlation functions of the Richardson pairing  
model
(also known as the reduced or discrete-state BCS model) in the
canonical ensemble.  Making use of the Algebraic Bethe Ansatz  
formalism, we obtain
exact expressions which are easily evaluated numerically for any value  
of the
pairing strength up to large numbers of particles.
We provide explicit results at half-filling and extensively
discuss their finite-size scaling behavior.
\end{abstract}

\maketitle

\section{Introduction}

The pairing phenomenon is ubiquitous in quantum many-body systems
of sizes ranging from the very small, like quarks and nuclei, to
the very large, like stars \cite{Rajagopalh0011333,AlfordARNPS51}.
The common feature of all these seemingly
unrelated systems is the instability against the formation of Cooper
pairs for an arbitrarily weak attractive force, the basis of the BCS  
theory
of superconductivity \cite{bcs-57}.
Despite the diverse nature of pairing systems, many of their fundamental
properties can be understood phenomenologically from a so-called  
reduced BCS model,
\be
H_{BCS}=\sum^N_{\stackrel {\a=1}{\sigma=+, -}}
\frac{\e_\a}2  c^\dagger_{\a\sigma}c_{\a\sigma}
  -g\sum^N_{\a,\b=1} c^\dagger_{\a+}c^\dagger_{\a-}
  c_{\b-}c_{\b+},\label{BCSH}
\ee
which was introduced by Richardson in the early 1960's in the context  
of nuclear physics
\cite{rs-62}.
The model simply describes (pseudo) spin-$1/2$ fermions
(electrons, nucleons, etc\dots) in a shell of doubly degenerate single
particle energy levels with energies $\e_\a/2$, $\a = 1, \dots N$.
$c_{\a,\s}$ are the
annihilation operators, $\s = +,-$ labels the degenerate time
reversed states (i.e. spin or isospin) and
$g$ denotes the pairing coupling constant.
Despite its simplified character (all levels interact uniformly),
this Hamiltonian captures the main essence of the problem;
fixing (ab-initio or phenomenologically) the energy levels $\e_\a$ and  
the
coupling $g$ allows to obtain quantitative predictions, since the model
remains solvable for an arbitrary choice of parameters.

In the thermodynamic limit, and within the grand-canonical ensemble,
the properties of the Richardson model are correctly described via the
BCS variational ansatz \cite{bcs-57}.
However, for finite numbers of particles, the situation is more complex.
The actual solution then depends on the ensemble chosen, and for  
physically relevant
systems the grand-canonical ensemble is not always the appropriate one.
For example, nuclei have a fixed number of nucleons;
due to the typically large charging energy, experiments on ultra-small
metallic grains are also performed at fixed number of electrons  
\cite{exp}.
In these cases a treatment based on the canonical ensemble would be
more appropriate, precluding a BCS mean-field approach.  Moreover,
dealing with a system in the mesoscopic regime precludes the use of
quantum statistical mechanics, and one is thus forced to rely either
on uncontrolled approximations, or nonperturbative methods.

Fortunately, Richardson's Hamiltonian (\ref{BCSH}) is one of the
theories for which an exact solution can be constructed in the canonical
ensemble \cite{rs-62}.
This solution explains several interesting features of the
mesoscopic physics of superconductors, complementing previous  
approximate
treatments (see the review [\onlinecite{dr-01}]).
In particular it allowed to give a definitive answer to Anderson's 1959
question \cite{a-59}: {\it What is the size limit for a metallic grain  
to
have superconducting properties?} The utility of the model is thus
indisputable (see also the reviews [\onlinecite{dps-04,dh-03}] for some  
non
condensed-matter applications), however most of the attention
has been concentrated on thermodynamical quantities.
On the other hand, experiments typically give access to
static or dynamical correlation functions, which are not easily obtained
in this framework. Richardson himself in 1965 \cite{r-65} derived a  
first
exact expression for static correlation functions, which unfortunately  
has a
degree of complexity that grows factorially with system size, and was
therefore not suitable for actual calculations.
In a significant development, Amico and Osterloh \cite{ao-02} proposed  
a new method (based
on a generalization of earlier work by Sklyanin \cite{sk}) to write  
down such
correlations explicitly.  The complexity of this method was still  
factorial and all the numerical
results were therefore limited to system sizes of up to 16 particles.
A disadvantage of these methods is that all the eigenstates
of the Hamiltonian must be known to get the correlation functions.

A major simplification was then proposed by Zhou et al. \cite{zlmg-02}  
(see also
[\onlinecite{lzmg-03}]).  Using the Algebraic Bethe Ansatz (ABA) and  
the Slavnov
formula for scalar products of states \cite{s-89}, they managed to  
write the static
correlation functions as sums over $N_r^2$ determinants of $N_r\times  
N_r$ matrices, reducing the complexity of
the problem to order $N_r^5$ (here $N_r$ is the number of rapidities in  
the eigenstates).  Furthermore, in this approach only
the knowledge of the ground-state wavefunction is required.
Surprisingly, this approach has not been used until now to obtain  
quantitative
numerical results for the correlation functions, with the notable
exception of the calculation of ground-state entanglement properties  
\cite{dlz-05}.

In this paper we fill this gap. As a first step we reanalyze the  
results of
Refs. [\onlinecite{zlmg-02,lzmg-03}], rewriting all static correlation
functions as sums over only $N_r$ determinants (thereby reducing the
complexity of the problem by a further factor of $N_r$). We then  
provide analytical
formulas for the physically relevant correlation functions,
and evaluate them for some model Hamiltonians.
We stress that having reduced the complexity of the problem by this  
amount,
correlation functions of systems with many more particles than before  
can be calculated on a simple computer.
In this way we can describe the crossover from mesoscopic to  
macroscopic physics, going
beyond previous results limited to fewer particles \cite{mff-98,ao-02}.

The paper is organized as follows.
In Sec. \ref{secmod} we discuss the model and its general properties.
In Sec. \ref{secaba} we recall how to calculate static correlation  
functions
by means of Algebraic Bethe Ansatz, and derive their general expressions
in terms of sums of $N_r$ determinants.
This section is rather technical, thus the reader interested in the
physical results can skip directly to
Sec. \ref{secgs} where we discuss how to solve the Richardson equations  
for
the ground state, and derive quantities that do not require
knowledge of the determinant representation.
In Sec. \ref{seccorr} all the
correlation functions are explicitely calculated at half-filling.
The paper is closed by Sec. \ref{concl} where we also
discuss open problems for future investigation.

\section{The Model}
\label{secmod}

A simple but very important property of the system is the
so-called blocking effect \cite{rs-62,dr-01}, i.e. unpaired particles
completely decouple from the dynamics and behave as if they were free.
We will denote the total number
of fermions as $N_f$, and the total number of pairs as $N_p$.
Due to level blocking, we will only consider $N_f=2 N_p$ paired  
particles in $N$
unblocked levels.  In terms of pair annihilation and creation operators
\be
b_\a= c_{\a-}c_{\a+}\, \qquad b_\a^\dagger=  
c^\dagger_{\a+}c^\dagger_{\a-}\,,
\ee
the Hamiltonian is
\be
H= \sum^N_{\a=1}
\e_\a  b^\dagger_\a b_\a -g\sum^N_{\a,\b=1} b^\dagger_\a b_\b\,,
\ee
and $n_\a= 2b_\a^\dagger b_\a$ is the number of
particles in level $\a$.

The pair creation and annihilation operators satisfy the commutation  
relations
\be
[b_\a,b_\b^\dagger]= \delta_{\a\b}(1-2b^\dagger_\a b_\a)\,,\qquad
[b_\a,b_\b]=[b_\a^\dagger,b_\b^\dagger]=0\,.
\ee
The term $2b^\dagger_\a b_\a$ in the first commutator makes the model  
different from free
bosons and therefore non-trivial.

Using the pseudo-spin realization of electron pairs
$S^z_\a= b^{\dagger}_\a b_\a -1/2$, $S^-_\a= b_\a$,  
$S^+_\a=b^\dagger_\a$,
the BCS Hamiltonian becomes (up to a constant)
\be
H=\sum^N_{\a=1}
\e_\a S^z_\a -g \sum^N_{\a,\b=1} S^+_\a S^-_\b \,.
\label{spinH}
\ee
The operators $S_\a^{\pm,z}$ obey a standard spin algebra and so
the Hamiltonian (\ref{spinH}) describes a spin-$1/2$ magnet with  
long-range
interaction for the $XY$ components in a site-dependent transverse  
magnetic field
$\e_\a$.
Such a magnetic Hamiltonian is known in the literature as a
Gaudin magnet \cite{g-book}.
An important relation is
\be
S^\pm_\a S^{\mp}_\a=S_\a^2-(S_\a^z)^2\pm S_\a^z\,.
\label{comm}
\ee

Since the normalization of the pairing strength in the literature is  
not uniform, care must be taken when comparing the results we will  
obtain with other
papers (e.g. our $g$ is the half of the coupling used in
Refs. [\onlinecite{ao-02,mff-98}]).

\subsection{Grand-canonical BCS wavefunction}

In the grand-canonical (GC) ensemble the ground-state wavefunction is  
the BCS
variational ansatz
\be
|GS\rangle=\prod_\a (u_\a +e^{i\phi_\a} v_\a b^\dagger_\a)|0\rangle\,,
\qquad {\rm with }\; u_\a^2+v_\a^2=1\,,
\ee
where the variational parameters $u_\a$ and $v_\a$ are real and  
$\phi_\a$ is
a phase which, it turns out, must be $\a$-independent.
$|GS\rangle$ is not an eigenstate of the particle number operator $N_f$
and the average condition $\langle N_f \rangle= \bar{N}_f$ determines
the GC chemical potential. Likewise, the commonly used definition
\be
\Delta_{GC}= 2 g \sum_\a \langle b_\a \rangle=
2 g \sum_\a u_\a v_\a e^{i\phi_\a}\,,
\label{DelGC}
\ee
for the superconducting gap makes sense only in a GC ensemble, since
$\langle b_\a \rangle$ is zero when evaluated at fixed particle number.
The variational parameters are obtained as
\be
v_\a^2=\frac12\left[1-\frac{\e_\a-\mu}{\sqrt{(\e_\a- 
\mu)^2+|\Delta_{GC}|^2}}
\right]\,,
\label{vjBCS}
\ee
where $\mu$ is the GC chemical potential.

It is then easy to calculate (static) correlation functions on this GS:
\be
\langle b_\a^\dagger b_\a \rangle= v_\a^2,\qquad
\langle b_\a b_\a^\dagger \rangle= u_\a^2,\qquad
\langle b_\a^\dagger b_\b \rangle= u_\a v_\a u_\b v_\b
\,.
\ee

\subsection{Canonical description and Richardson solution}

The exact solution of (\ref{BCSH}) was worked out by Richardson  
\cite{rs-62}.
In the canonical ensemble the model is
integrable \cite{crs-97} and tractable by means of algebraic
methods \cite{aff-01,dp-02,zlmg-02,lzmg-03}.
We review here only the main points of this solution.

In the ABA, eigenstates are contructed by applying raising operators on
a so-called reference state (pseudovacuum).  We here choose the  
pseudovacuum
(in the spin representation) to be fully polarized along the $\hat{z}$  
axis
  \be
S_\a^z |0\rangle = \frac{1}{2} |0\rangle\,, \quad \forall \ \a.
\label{pv}
\ee
In the pair representation, this state thus corresponds to having one  
pair in each available level.
Eigenstates with $N_p$ pairs are then characterized by $N_r = N-N_p$  
spectral parameters
(rapidities) $w_j$, and take the form of Bethe wavefunctions
\be
|\{w_j\}\rangle=\prod_{k=1}^{N_r} \mathcal{B}(w_k) |0\rangle\,.
\label{RICHWF}
\ee
The operators $\mathcal{B}$, together with operators $\mathcal{A},  
\mathcal{C}, \mathcal{D}$
defined as
\bea
&&\!\!\mathcal{A}(w_k) = \frac{-1}{g} + \sum_{\alpha=1}^N  
\frac{S^z_\a}{w_k - \e_\a}, \hspace{0.2cm}
\mathcal{B}(w_k)=\sum_{\a=1}^N \frac{S_\a^-}{w_k - \e_\a}, \nonumber \\
&&\!\!\mathcal{C}(w_k)=\sum_{\a=1}^N \frac{S_\a^+}{w_k - \e_\a},  
\hspace{0.2cm}
\mathcal{D}(w_k) = \frac{1}{g} - \sum_{\alpha=1}^N \frac{S^z_\a}{w_k -  
\e_\a}
\label{inverse_problem}
\eea
obey the Gaudin algebra, which is the quasi-classical limit of the  
quadratic Yang-Baxter
algebra associated to the $gl(2)$ invariant $R$-matrix (we refer the  
readers to [\onlinecite{lzmg-03}]
for details).

The wavefunctions (\ref{RICHWF}) are eigenstates of the transfer matrix,
and thus of the Hamiltonian (\ref{BCSH}), when the parameters $w_j$  
satisfy
the Richardson equations
\be
\frac1g=\sum_{\a=1}^N\frac1{w_j - \e_\a}-\sum_{k\neq j}^{N_r}  
\frac2{w_j-w_k}\,\quad
j=1,\dots, N_r\,.
\label{RICHEQ}
\ee
Throughout the paper we will refer with latin indices to the rapidities
and with greek ones to the energy levels.
The total energy of a Bethe state is
$E= \sum_{\alpha=1}^N\frac{\epsilon_\a}{2}- \sum_j w_j + g(2N_r-N)$.  
For a
given $N$ and $N_r$ the number of solutions of Richardson equations is
$\binom{N}{N_r}$, and coincides with the
dimension of the Hilbert space of $N_r$ pair vacancies distributed into  
$N$
different levels, i.e. the solutions to Richardson equations give
all the eigenstates of the model.

Note that the Richardson equations (\ref{RICHEQ}) have a different sign  
of $g$
compared to the ones mostly considered in the literature. This is due  
to the
particular choice of the pseudovacuum we made in Eq. (\ref{pv}),  
whereas the
most common choice is $S_\a^z |0\rangle= -1/2 |0\rangle$.
With our choice of pseudovacuum, Bethe states are built by destroying  
pairs, as
in Eq. (\ref{RICHWF}) and not by creating them. We use this somehow  
unusual
pseudovacuum following Refs. [\onlinecite{zlmg-02,lzmg-03}] in order to  
use all the
formulas there without any adaptation. At half-filling (that is the  
only case
considered numerically here) the different choice of the pseudovacuum  
only
matters as a global normalization and a different labeling of the  
states.

The connection between the canonical and grand-canonical ensembles was  
first pointed out by
Richardson himself \cite{r-77}, who showed how in the large $N_f$ limit
one recovers the BCS gap equation as
\be
N_f=\sum_{\a=1}^N
\left(1-\frac{\e_\a-\mu}{\sqrt{(\e_\a-\mu)^2+|N_f\Delta|^2}}\right)\,,
\label{Deltagen}
\ee
where now $\mu$ is fixed by the density and the equation can be solved  
to find
$\Delta$, which with this normalization is an {\it intensive} quantity  
and
corresponds to $\Delta_{GC}/N_f$.
For the ground-state energy per pair $E_0$ one finds
\be
N_p E_0= \sum_{\a=1}^N \e_\a\left(
   1-\frac{\e_\a-\mu}{\sqrt{(\e_\a-\mu)^2+ |N_f\Delta|^2}}\right) -
\frac{N_f^2\Delta^2}{2g}\,.
\label{E0rich}
\ee
Anderson \cite{a-59} argued that increasing the mean energy spacing $d$  
(that
is inversely proportional to the volume in a metallic grain)  
superconductivity
should disappear when $d$ becomes of the order of the bulk gap  
$\Delta_{GC}$.
Our study of correlation functions to be presented below clearly shows  
this crossover.

\section{Algebraic Bethe Ansatz and correlation functions}
\label{secaba}

The starting point to calculate correlation functions with the  
Algebraic Bethe
Ansatz (ABA) is having a representation for the scalar products
of two generic states defined by $N_r$ rapidities ($N-N_r$ pairs)
\be
\langle \{w\}|\{v\}\rangle=
\langle 0| \prod^{N_r}_{b=1} \mathcal{C}(w_b) \prod^{N_r}_{a=1}  
\mathcal{B}(v_a) |0\rangle\,,
\ee
when at least one set of parameters (e.g. $w_b$ but not $v_a$)
is a solution to the Richardson equations.  Following standard  
notations, $\mathcal{C}$ is the conjugate of the operator  
$\mathcal{B}$.
Such a representation exists, and is known as the Slavnov formula  
\cite{s-89},
which for the case at hand specifically reads \cite{zlmg-02}
\bea
\langle \{w\}|\{v\}\rangle &=&
\frac {\prod^{N_r}_{a \neq b}(v_b-w_a)}
{\prod _{b <a} (w_b -w_a) \prod _{a <b} (v_b -v_a)}
\nonumber\\ && \times {\rm det}_{N_r} J(\{ v_a \}, \{ w_b \})\,,
\eea
where the matrix elements of $J$ are given by
\bea
J_{ab} &=& \frac {v_b -w_b}{v_a -w_b}
\left ( \sum^N_{\a=1} \frac {1}
{(v_a -\e _\a)(w_b -\e_\a)}\right. \nonumber\\ && \left.-2\sum _{c\neq  
a}^{N_r} \frac {1}{(v_a-v_c)(w_b -v_c)}
\right ).
\eea
from which the norms of states simply follow from $v\to w$ as
$ ||\{v\}||^2=\det_{N_r} G$ with a Gaudin matrix
\be
G_{ab}=
\begin{cases}\displaystyle
\sum_{\b=1}^N \frac1{(v_a-\e_\b)^2}-2\sum_{c\neq  
a}^{N_r}\frac1{(v_a-v_c)^2}\quad &
a=b\,,\\ \displaystyle
\frac2{(v_a-v_b)^2}& a\neq b\,,
\end{cases}
\label{Gaudin}
\ee
recovering Richardson's old result \cite{r-65}.

The key point is that any form factor of a local spin operator between  
two Bethe eigenstates can
be represented via (\ref{inverse_problem}) as a scalar product with
one set, e.g.  $\{v\}$ not satisfying the Bethe equations, for which
Slavnov's formula is applicable.
This has been explicitly worked out in Ref. [\onlinecite{zlmg-02}].
For $\{w\}$,$ \{v\}$ containing respectively $N_r+1$ and $N_r$  
elements, the
nonzero form factors are:
\bea
\langle \{w\}|S^-_\a|\{v\}\rangle =
  \langle \{v\}|S^+_\a|\{w\}\rangle =
\nonumber\\ \frac {\prod^{N_r+1}_{b=1} (w_b - \e_\a)} {\prod  
^{N_r}_{a=1} (v_a - \e_\a)}
  \frac { {\rm det}_{N_r +1} T (\a, \{w\}, \{ v\})}
{\prod _{b > a} (w_b -w_a) \prod _{b <a} (v_b -v_a)}\,,
\label{matels-}
\eea
and, for both $\{w\}$ and $\{v\}$ containing $N_r$ rapidities
\bea
\langle \{w\}|S^z_\a|\{v\}\rangle =
\prod^{N_r}_{a=1} \frac {(w_a - \e_\a)} {(v_a - \e_\a)} \nonumber\\  
\times
\frac { {\rm det}_{N_r} \left (\frac12 T_z(\{w\}, \{v\})
- Q (\a, \{ w\}, \{ v\}) \right )}
{\prod _{b > a} (w_b -w_a) \prod _{b <a} (v_b -v_a)}\,,
\eea
with the matrix elements of $T$ given by
\bea
T_{ab}(\a) =&&
\prod ^{N_r+1}_{{c \neq a}} (w_c - v_b)
\left ( \sum^N_{\a=1} \frac1{(v_b -\e _\a)(w_a -\e_\a)}
\right.\nonumber\\ && \left.-2\sum _{c \neq a} \frac1{(v_b-w_c)(w_a  
-w_c)} \right )\,,
~~b < N_r+1, \nonumber\\
T_{a N_r+1}(\a)  && =  \frac {1}{(w_a -\e_\a)^2}, \ \
Q_{ab}(\a) = \frac {\prod _{c \neq b} (v_c-v_b)} {(w_a-\e_\a)^2}.
\nonumber\\
\eea
Above, $T_z$ is the $N_r \times N_r$ matrix
obtained from $T$ by deleting the last row and column and replacing  
$N_r+1$ by
$N_r$ in the matrix elements. Here it is assumed that both $\{ v_a \}$  
and
$\{ w_b \}$ are solutions to Richardson's Bethe equations.
However, the results are still valid for $S^\pm_\a$ if only
$\{ w_b \}$ satisfy the Bethe equations.

\subsection{Determinant representation of the correlation functions}

In Ref. [\onlinecite{zlmg-02}] it has been pointed out that due to the  
simplicity
of the solution of the ABA not only the form factors, but any static  
correlation
function can be written in a determinant representation.
This simplicity puts the BCS model in an extremely privileged position  
for
a detailed study of the static correlation functions.

The result for $\langle\{w\}|S^-_\a S^+_\b|\{v\}\rangle$ has been  
explicitly
worked out \cite{zlmg-02}
\begin{multline}
\langle \{w\}|S^-_\a S^+_\b|\{v\}\rangle =
\sum _{i=1}^{N_r} \frac1{v_{i} -\e_\b}
\langle \{w\}|S^-_\a|\{v\}_i\rangle \\ -
\sum _{i' \neq i} \frac {1}{(v_i -\e_\b)(v_{i'} -\e_\b)}
\langle \{w\}|S^-_\a S^-_\b|\{v\}_{i,i'}\rangle.
\label{matels-s+}
\end{multline}
Here the sets indicated by $\{v\}_i$ stands for sets where
the rapidity $i$ has been removed and similarly for
$\{v\}_{i,i'}$ both $i$ and $i'$ rapidities have been
removed.
The $S^-S^-$ form factor is given by \cite{zlmg-02}
\bea
\langle \{w\}|S^-_\a S^-_\b|\{v\}\rangle =
\frac {\prod ^{N_r}_{b=1} (w_b - \e_\a)(w_b-\e_\b)}
{\prod ^{N_r-2}_{a=1} (v_a - \e_\a)(v_a -\e_\b)}
\nonumber\\ \times \frac { {\rm det}_{N_r} T (\a,\b, \{ w_b \}, \{ v_a  
\})}
{\prod _{b > a} (w_b -w_a) \prod _{b <a} (v_b -v_a)}\,,
\label{matels-s-}
\eea
with
\bea
&&T_{ab}(\a,b) =
\prod^{N_r}_{c \neq a} (w_c - v_b) \left( \sum^N_{\gamma=1}
\frac{1}{(v_b -\e_\gamma)(w_a -\e_\gamma)}\right.\nonumber\\
&&-2\left.\sum _{c \neq a}
\frac{1}{(v_b-w_c)(w_a-w_c)} \right),~~~b < N_r-1, \nonumber\\
&&T_{aN_r-1}(\a,\b)  =
\frac {2w_a-\e_\a-\e_\b}{[(w_a -\e_\a)(w_a-\e_\b)]^2},\nonumber\\
  &&T_{aN_r}(\a,\b)  = \frac {1}{(w_a -\e_\a)^2}\,,
\eea
where $\a\neq\b$ is assumed, with the convention
that it vanishes when $\a=\b$.
Note that $\langle \{w\}|S^-_\a S^-_\b|\{v\}\rangle$ is symmetric under  
the
exchange of $\a$ and $\b$, although this is not manifest in the formal
expression.  This nontrivial property will be checked during the  
numerical
computation.

This correlation is then written as the sum of $N_r^2$ determinants,
which is much less than the sum over the full Hilbert space needed in  
other
approaches.  In the following we will determine a similar expression for
$\langle S_\a^z S_\b^z\rangle$ and then we will show that it is possible
to reduce these formulas to sums of only $N_r$ terms.

\subsubsection{Determinant representation of $\langle S_\a^z  
S_\b^z\rangle$}

The operator $\mathcal{A}(u)$ only has simple poles
at $\e_\a$ such that \cite{lzmg-03}
\be
S^z_\a = \lim_{u \to \e_\a}(u-\e_\a) \mathcal{A}(u)\,.
\ee
This allows one to write
\begin{multline}
\bra{\{w\}} S^z_\a S^z_\b \ket{\{v\}} = \lim_{u \to \e_\b} \lim_{u' \to  
\e_\a}
  (u'-\e_\a)(u-\e_\b)
\\ \times
\bra{\{w\}}\mathcal{A}(u') \mathcal{A}(u)\prod_{i=1}^{N_r}  
\mathcal{B}(v_i)\ket{0}
\label{limdef}\,,
\end{multline}
which can be easily computed by commuting the $\mathcal{A}$ operators  
until
they reach the far right and act on the pseudovacuum in the following  
way:
\be
  \mathcal{A}(u)\ket{0} = a(u)\ket{0} = -\frac{1}{g} \ket{0}
+ \frac{1}{2}\sum_{\gamma=1}^{N}\frac{1}{u-\e_\gamma} \ket{0}.
  \label{actvac}
\ee
Using the commutation relation \cite{lzmg-03}
\be
\left[\mathcal{A}(u), \mathcal{B}(v)\right] =  
\frac{\mathcal{B}(u)}{u-v} -
\frac{\mathcal{B}(v)}{u-v}\,
\ee
and defining $\mathcal{G}\equiv \bra{\{w\}}\mathcal{A}(u')  
\mathcal{A}(u)\prod_{i=1}^{N_r} \mathcal{B}(v_i)\ket{0}$,
we find by commuting $\mathcal{A}(u)$ and $\mathcal{B}(v_1)$ that
\begin{multline}
\mathcal{G}=
\bra{\{w\}}\mathcal{A}(u') \mathcal{B}(u)(v_1)\mathcal{A}(u)
\prod_{i=2}^{N_r} \mathcal{B}(u)(v_i)\ket{0} \\ +
\frac{\bra{\{w\}}\mathcal{A}(u') \ket{\{v\}_1 ; u }}{u-v_1}
- \frac{\bra{\{w\}}\mathcal{A}(u') \ket{\{v\}}}{u-v_1}\,,
\end{multline}
where $\ket{\{v\}_1 ; u }$ is the (non-Bethe) state built by replacing  
the
rapidity $v_1$ by $u$.
By commuting again $N_r-1$ times and using Eq. (\ref{actvac}), we find
\bea
\mathcal{G} &=&
F(u) \bra{\{w\}}\mathcal{A}(u') \ket{\{v\}}
\nonumber\\ &&+
\sum_{i=1}^{N_r}
\frac{1}{u-v_i}
\bra{\{w\}}\mathcal{A}(u') \ket{\{v\}_i ; u }\,,
\eea
where we defined
$F(u)\equiv-\frac1g+ \frac12\sum_{\gamma=1}^N\frac1{u-\e_\gamma} -
\sum_{i=1}^{N_r} \frac{1}{u-v_i}$.
The same procedure can then be repeated in
order to have $\mathcal{A}(u')$ act on $\ket{0}$:
\bea
\mathcal{G}
&=&
F(u)
\left[
F(u')
\vm{\{w\}|\{v\}}
+
\sum_{i=1}^{N_r} \frac{\vm{\{w\}|\{v\}_i ; u' }
}{u'-v_i}
\right]
\nonumber\\ && +
\sum_{i=1}^{N_r}
\frac{1}{u-v_i}
\left[ \vphantom{\frac{1}{u'-v_{i'}}}
\sum_{i' \ne i}^{N_r} \frac{1}{u'-v_{i'}}
\vm{\{w\}|\{v\}_{i,i'} ; u,u'}
\right. \nonumber\\ && \left.
+
F_{i}(u')
\vm{\{w\}|\{v\}_i ; u}
+
\frac{1}{u'-u}
\vm{\{w\}|\{v\}_{i} ; u'}
\right. \nonumber\\ && \left.
-
\frac{1}{u'-u} \vm{\{w\}|\{v\}_i ; u}
\right]\,,
\eea
with $F_{i}(u) \equiv -\frac1g  
+\frac12\sum_{\gamma=1}^N\frac1{u'-\e_\gamma}
-\sum_{i' \ne i}^{N_r} \frac{1}{u'-v_{i'}}$.
It is then easy to take the limit as prescribed by Eq. (\ref{limdef})  
and find
\begin{multline}
\bra{\{w\}} S^z_\a S^z_\b \ket{\{v\}} = \frac{\vm{\{w\}|\{v\}}}{4}
+\sum_{i=1}^{N_r} \frac{\bra{\{w\}} S^-_\a\ket{\{v\}_i}}{2(\e_\a-v_i)}
\\ + \sum_{i=1}^{N_r} \frac{\bra{\{w\}}  
S^-_\b\ket{\{v\}_i}}{2(\e_\b-v_i)}+
\sum_{i' \ne i}^{N_r} \frac{\bra{\{w\}} S^-_\a S^-_\b\ket{\{v\}_{i,i'}}
}{(\e_\b-v_{i})(\e_\a-v_{i'})}\,.
\label{szsz}
\end{multline}
Note that we use the same notation as in Eq. (\ref{matels-}) for the  
$S^-$
form factor but there it is evaluated between two states with $N_r$ and  
$N_r+1$
rapidities, while here it is between two states with $N_r-1$ and $N_r$.
This should not be a source of confusion.

The static correlation function can now be evaluated by setting
$\{v\}=\{w\}=\{w_0\}$; the set of rapidities corresponding to the  
ground state
of the system. Using equations (\ref{matels-}) and (\ref{matels-s-}),  
we can
directly express this correlation function as a sum of $N_r^2+N_r$   
determinants.

\subsection{Reduction formulas}

As anticipated in the introduction, the previous expressions can be  
reduced
to sums over only $N_r$ determinants. This is explicitely worked out in  
the following.
We will assume that $\a\neq \b$, because for intra-level
correlations from Eq. (\ref{comm}) we have
\be
\bra{\{w\}} S^-_\a S^+_\a\ket{\{w\}} =\frac12+\bra{\{w\}} S^z_\a  
\ket{\{w\}}\,,
\label{odonlevel}
\ee
which is already a single determinant expression.

\subsubsection{Reduction of $\langle S_\a^- S_\b^+\rangle$}
We here need to evaluate
Eq. (\ref{matels-s+}) in the limit $v\to w$. In this case, Eqs.  
(\ref{matels-})
and (\ref{matels-s-}) for the form factors simplify to
\be
\frac{\langle \{w\}_{N_r} |S_\a^- |\{w_q\}_{N_r-1}\rangle}{w_q-\e_\b}=
\frac{w_q-\e_\a}{w_q-\e_\b} \mbox{det}_{N_r} U^{(q)}\,,
\label{SZcont}
\ee
\begin{multline}
\frac{\langle\{w\}_{N_r} |S_\a^-S_\b^-|\{w_{q,l}\}_{N_r-2}\rangle}{
(w_q-\e_\b)(w_l-\e_\b)}=
-\frac{(w_q-\e_\a)(w_l-\e_\a)}{w_l-w_q} \\
\times \mbox{det}_{N_r} U^{(ql)}\,.
\label{S+S+}
\end{multline}
The matrices $U$ are defined as follows.  $U^{(q)}$ and $U^{(q,l)}$ are  
equal to the Gaudin
matrix (\ref{Gaudin}) except for columns $q$ and $q,l$ respectively,  
where
\bea
U^{(q)}_{aq}&=& U^{(ql)}_{aq}=\frac1{(w_a-\e_\a)^2}\,, \label{vecC}\\
U^{(ql)}_{al}&=& \frac{2w_a-\e_\a-\e_\b}{(w_a-\e_\a)^2(w_a-\e_\b)^2}\,.
\label{vecB}
\eea
In Eq. (\ref{S+S+}) it is explicitely assumed that $\a\neq\b$ with the
convention that for $\a=\b$ it is zero. And in fact, for $\a=\b$ it is
not difficult to prove that Eq. (\ref{SZcont}) reproduces the correct  
result
given by Eq. (\ref{odonlevel}).

Since everything is symmetric under exchange of $l$ and $q$ we only  
perform
the sum over $l<q$ and in the end we multiply the result by 2.
Thus we need to perform the sum (neglecting for the moment the $l$  
independent factors)
\be
\sum_{l=1}^{q-1} \frac{w_l-\e_\a}{w_l-w_q} \det U^{(ql)}\equiv
\sum_{l=1}^{q-1} K_{lq} \det U^{(ql)}\,.
\label{sum}
\ee
Let us write the matrix $U^{(ql)}$ as a vector of vectors
\be
U^{(ql)}=|\vec{G}_1\dots \vec{G}_{l-1}, \vec{B}, \vec{G}_{l+1}\dots
\vec{G}_{q-1}, \vec{C}, \vec{G}_{q+1}\dots \vec{G}_{N_r}|\,,
\ee
where the $\vec{G}_i$ corresponds to the columns of the matrix $G$,
$\vec{C}$ (at position $q$) corresponds to the vector given by
Eq. (\ref{vecC}) and $\vec{B}$ by Eq. (\ref{vecB}).

The sum we want to calculate is (we use $| \cdot |$ for the determinant)
\begin{multline}
\sum_{l=1}^{q-1}K_{lq} \det U^{(ql)} = K_{1q}  
|\vec{B},\vec{G}_2,\vec{G}_3 \dots|
  \\  +
K_{2q} |\vec{G}_1, \vec{B}, \vec{G}_3 \dots|+
K_{3q} |\vec{G}_1, \vec{G}_2, \vec{B} \dots|+
\dots
\label{suml}
\end{multline}
Using the fact that two determinants which differ by a single column can
easily be expressed as a single determinant the two first terms of the  
sum can
be written as $|K_{2q}\vec{G}_1- K_{1q}\vec{G}_2,  \vec{B}, \vec{G}_3  
\dots|$.
Elementary column operations allow us to write the third determinant as
$K_{3q} |\vec{G}_1, \vec{G}_2, \vec{B} \dots| =  K_{3q} |\vec{G}_1-
\frac{K_{1q}}{K_{2q}} \vec{G}_2, \vec{G}_2, \vec{B} \dots|$. This term  
differs
by a single column from the preceding sum. We can then write the sum of  
the
first three terms as a single determinant
$K_{3q}|\vec{G}_1-\frac{K_{1q}}{K_{2q}} \vec{G}_2,
\vec{G}_2- \frac{K_{2q}}{K_{3q}} \vec{G}_3, \vec{B} \dots|$.
We can keep on adding terms in the same way until we reach
column $q-1$ and find that Eq. (\ref{suml}) can be written as a single
determinant
\be
K_{q-1 q}|\vec{G}_1- \frac{K_{1q}}{K_{2q}} \vec{G}_2,
\vec{G}_2- \frac{K_{2q}}{K_{3q}} \vec{G}_3, \dots \vec{B}, \vec{C},
\vec{G}_{q+1}\dots \vec{G}_{N_r}|\,.
\ee

In this way we reduced the double sum to a single one. The additional  
terms in
the correlation function (coming from $\vm{S^-}$) can also be  
incorporated to
the sum in a similar fashion.
The $\vm{S^-}$ term in Eq. (\ref{matels-s+}) is given by Eq.  
(\ref{SZcont})
and can be simply encoded in the representation we just obtained
for the sum over $l<q$ of $\vm{S^-S^-}$.
In this way, we finally have for the full correlation function
\be
\langle \{w\} |S_\a^- S_\b^+ |\{w\}\rangle= \sum_{q=1}^{N_r}
\frac{w_q-\e_\a}{w_q-\e_\b} D_q^{(\a,\b)} \,,
\label{S+S-red}
\ee
where we defined the matrix
\be D_q^{(\a,\b)} =
\left[\vec{D}_{q,1}^{(\a,\b)},\vec{D}_{q,2}^{(\a,\b)},\dots
\vec{D}_{q,N_r}^{(\a,\b)}
\right]
\label{Dqmn}
\ee
that has the following structure
\be
D_{q,i}^{(\a,\b)} =
\begin{cases}\displaystyle
\vec{G}_i- \frac{K_{iq}}{K_{i+1q}} \vec{G}_{i+1} \ &
i<q-1,\\ \displaystyle
\vec{G}_{i}+2\frac{ (w_q-\e_\b) (w_{q-1}-\e_\a)}{w_{q-1}-w_q} \vec{B} \  
&
i=q-1,\\
\vec{C} \  &
i=q,
\\ \vec{G}_i& i> q.
\end{cases}
\ee
The low level of complexity of this representation as sum of $N_r$  
determinants
of $N_r$ by $N_r$ matrices allows us to access easily the static  
correlation
functions for systems with large number of pairs compared to previously
published results.

We stress again that these formulas are true only for $\a\neq \b$.

\subsubsection{Reduction of $\langle S_\a^z S_\b^z\rangle$}

According to Eq. (\ref{szsz}), when $v\to w$
\begin{multline}
\langle \{w\} |S_\a^z S_\b^z |\{w\}\rangle=
\frac14 \langle \{w\} |\{w\}\rangle \\
-\frac12\left[
\sum_{i=1}^{N_r} \frac{\langle \{w\} |S_\b^- |\{w\}_i  
\rangle}{w_i-\e_\b}-
\sum_{i'\neq i}^{N_r} \frac{\langle \{w\}  
|S_\a^-S_\b^-|\{w\}_{i,i'}\rangle}{(w_i-\e_\b)(w_{i'}-\e_\a)}
\right]\\
-\frac12\left[
\sum_{i=1}^{N_r} \frac{\langle \{w\} |S_\a^-  
|\{w\}_i\rangle}{w_i-\e_\a}-
\sum_{i'\neq i}^{N_r} \frac{\langle \{w\}  
|S_\b^-S_\a^-|\{w\}_{i,i'}\rangle}{(w_i-\e_\b)(w_{i'}-\e_\a)}
\right]\,.
\end{multline}
 >From Eqs. (\ref{S+S-red}) and (\ref{Dqmn}) it is straightforward to  
show that
when $\a\neq\b$
\begin{multline}
\langle \{w\} |S_\a^z S_\b^z |\{w\}\rangle =  \\
  \frac{||w||^2}{4} -
\frac12 \sum_{q=1}^{N_r} (\det D_q^{(\a,\b)}+\det D_q^{(\b,\a)})\,.
\end{multline}
For $\a=\b$ the result is trivially
$\langle \{w\} |(S_\a^z)^2|\{w\}\rangle =1/4$.

This completes the representation of the static correlation functions  
in terms
of determinants.  To make further progress, we need explicit results  
for the
ground-state rapidities $\{ w \}$, {\it i.e.} the lowest-energy  
solutions
to the Richardson equations.  The following section is devoted to this.

\section{The solution of the Richardson equations for the ground state}
\label{secgs}

\subsection{General properties}

\begin{figure*}[bth]
\begin{tabular}{ll}
  \includegraphics[width=7.5cm]{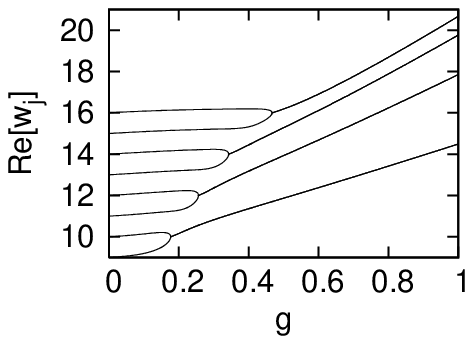} &
\includegraphics[width=7.5cm]{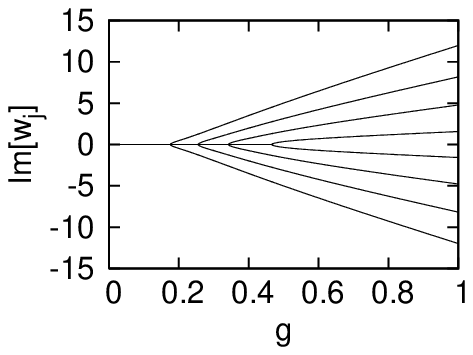} \vspace{3mm}\\
  \includegraphics[width=7.5cm]{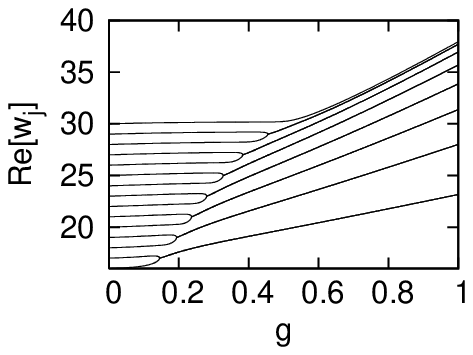} &
\includegraphics[width=7.5cm]{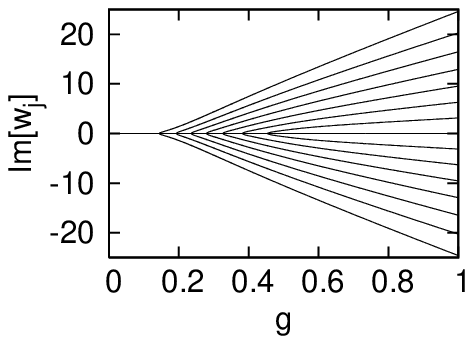} \vspace{3mm}\\
  \includegraphics[width=7.5cm]{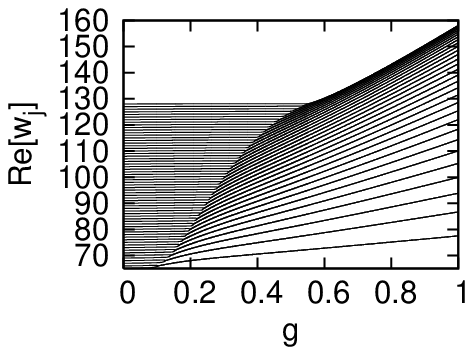} &
\includegraphics[width=7.5cm]{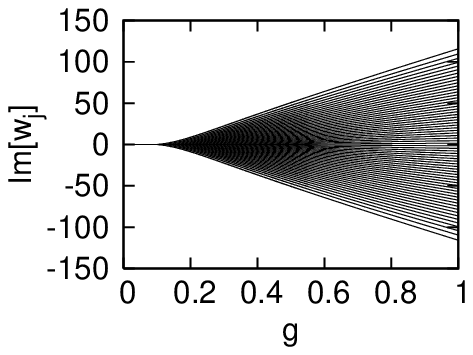} \vspace{3mm}
\end{tabular}
\caption{$g$ dependence of the real (left) and imaginary parts (right)
of the ground state rapidities. From top to bottom they correspond
to $N_r=8,15,64$ always at half-filling ($N=2N_r$).}
\label{GSPLOT}
\end{figure*}

At $g=0$, for $N_r$ pair vacancies in $N$ energy levels $\e_\a$ with  
only
double degeneracy, the $\binom{N}{N_r}$ solutions
to the Richardson equations are trivial. They are given by
Eq. (\ref{RICHWF}) with the $N_r$ rapidities set to be strictly equal  
to one of
the energies $\e_\a$. Clearly, the GS in that limit is built by
choosing the $N_r$ highest energy levels, i.e.
$w_1 = \e_N\,, w_2 =\e_{N-1} \ \dots \ w_{N_r} = \e_{N-N_r+1}$.

Apart from a few particular cases with a small number of particles, the
Richardson equations are not solvable analytically when $g\neq 0$.
The perturbative expansions for small\cite{r-66,sild-01}
and large \cite{yba-03,sg-06}
coupling are not predictive for all values of the pairing strength and  
so the
most accurate results come from the numerical solution.
The solutions are such that every $w_j$ is
either a real quantity or forms, with another parameter
$w_{j'}$, a complex conjugate pair (CCP), i.e. $w_{j'}^* =w_j$.
The mechanism for the CCPs formation is very easy:
as interactions are turned on, all $w_j$ are real quantities for small  
enough
$g$, but at a certain critical value of the coupling $g^*_j$
two rapidities will be exactly equal to one of the energy levels  
($w_j=w_{j'} =
\e_{\gamma}(j)$) and for $g> g^*_j$, the two parameters that collapsed
will form a CCP at least for a finite interval in $g$.  The situation is
in fact rather intricate:  the values $g^*_j$ are implicit functions of
all other rapidities, and can only be read off a full solution of the  
Richardson
equations for a specific choice of state.  Moreover, CCPs can split  
back into
real pairs, whose components can then re-pair with neighbouring  
rapidities.
Finding complex solutions to the Richardson equations is thus difficult  
in
general, since there is no equivalent to the 'string hypothesis' as
for {\it e.g.} integrable spin chains.

The solutions for the ground state have a particularly simple structure.
In fact, the set of critical points is such that the smaller a
rapidity is at $g=0$ the smaller the $g$ at which it forms a CCP will  
be.
As we raise $g$ from zero there will come a point at which $w_{N_r}$  
will form a
CCP with $w_{N_r-1}$ when they are both equal to $\e_{N-N_r+2}$.
As $g$ is raised some more, $w_{N_r-2}$ and $w_{N_r-3}$ will form a CCP  
at
$\e_{N-N_r+4}$ and this will go on until every rapidity has formed a  
CCP in the case
of even $N$.
Oppositely, with an odd number of rapidities in the system,
$w_1$ (the largest one at $g=0$) will always remain a real quantity no  
matter how
large the coupling strength is.
After the CCPs are formed no further collapse happens in the case of  
the ground
state, while for excited states further collapses can take place and  
complex
solutions can become real again.

Different choices of the parameters $\e_\a$ and of their eventual
degenerations specify different models.
In all the preceding sections everything was completely general
(modulo having to take some extra precautions in the case of coinciding  
levels $\e_\a$), but from now on we
specialize to the case of equally spaced doubly degenerate levels.
We make the choice to use $\e_\a =\a$ which sets the zero of energy and
implies that every energy will be given in units of the (pair)  
inter-level spacing.
Furthermore we consider only half-filling of the energy levels
$(N=2N_r = 2N_p =N_f)$.
In this case, as $g \to \infty$, the real part
of every rapidity will go to $+\infty$ whereas the CCPs imaginary parts  
will
go to $\pm\infty$.

\subsection{Numerical procedure and results}

At the precise value of $g$ at which a pair of rapidities $(w_j,w_{j'})$
collapse into a CCP $(w_j=w_{j'}=\e_\gamma(j))$, the Richardson  
equations (Eq. (\ref{RICHEQ})) labelled $j$ and $j'$ will include two  
diverging terms whose sum remains finite. In order to be able to treat  
these points numerically, one can define the following real variables,
\bea
w_{1,j} &\equiv&  w_j + w_{j'}
\\
w_{2,j} &\equiv&
\frac{ 2\e_{\gamma}(j) - w_j - w_{j'}}{\left(w_j-w_{j'}\right)^2},
\eea
whose inverse transformation reads
\bea
w_{j} &=& \frac{1}{2}
\left[w_{1,j}+  \sqrt{\frac{ 2\e_{j-1} - w_{1,j}}{w_{2,j}}} \right]\,,
\\
w_{j'} &=&  \frac{1}{2}
\left[w_{1,j} - \sqrt{\frac{ 2\e_{j-1} - w_{1,j}}{w_{2,j}}}\right] .
\eea
As discussed in Ref. [\onlinecite{r-66}],
we need to know beforehand which rapidities will
form a CCP and at which $\e_\gamma(j)$ it will happen in order to use
this type of change of variables. Since in this article we only need  
ground
state solutions, this requirement is easily met.

At the critical point $w_{2,j}$ goes to a well defined (though a priori
unknown) finite ${0}/{0}$ form. Using it as a variable in the system of
equations therefore avoids some potential numerical complications when  
close
to a critical point.

\begin{figure}[t]
  \includegraphics[width=8.5cm]{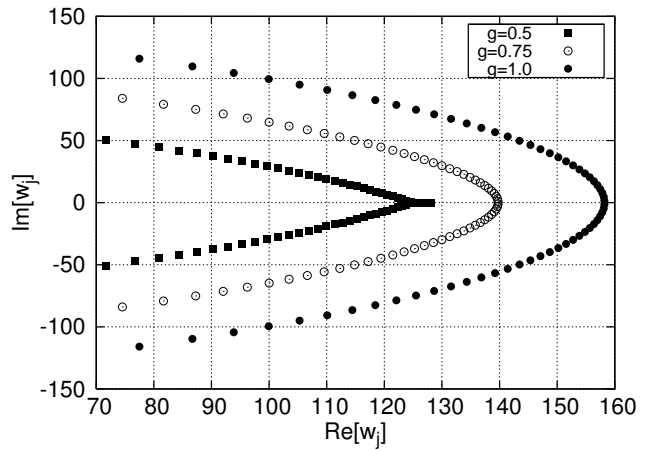}
\caption{Location in the complex plane of the ground-state rapidities
showing the formation of the arc-like solution. All the values  
correspond to
$N=128, N_r=64$.}
\label{arc}
\end{figure}

By multiplying the $j$ and $j'$ Richardson equations respectively by
$\e_{\gamma}(j)-w_{j}$ and $\e_{\gamma}(j)-w_{j'}$, we can get
rid of the divergences that show up at critical points. Adding the  
resulting
equations (giving $F_{1,j}$) and subtracting them and then dividing it  
by
$w_j-w_{j'}$ (giving $F_{2,j}$), it is simple to obtain the following  
two real
equations
\begin{multline}
F_{1,j} = \sum_{i\ne j-1}^{N}
\frac{(\e_{j-1}-\e_{i})(2\e_{i}-w_{1,j})}{(\e_{i})^2 -  
(\e_{i}-\frac{w_{1,j} }{4})w_{1,j}  - \frac{2\e_{j-1}-w_{1,j}}{4  
w_{2,j}}
}
\\
- \sum_{j'\ne j,j-1}^{N_r} 2  
\frac{(\e_{j-1}-w_{j'})(2w_{j'}-w_{1,j})}{(w_{j'})^2 -  
(w_{j'}-\frac{w_{1,j} }{4})w_{1,j}  - \frac{2\e_{j-1}-w_{1,j}}{4  
w_{2,j}}
}
\\
+\frac{2\e_{j-1}-w_{1,j} }{g}+2(N-1)-4(N_r-2) = 0\,,
\end{multline}
\begin{multline}
F_{2,j} = \sum_{i\ne j-1}^{N} \frac{(\e_{j-1}-\e_{i})}{(\e_{i})^2 -  
(\e_{i}-\frac{w_{1,j} }{4})w_{1,j}  - \frac{2\e_{j-1}-w_{1,j}}{4  
w_{2,j}}  }
\\
- \sum_{j'\ne j,j-1}^{N_r} 2 \frac{(\e_{j-1}-w_{j'})}{(w_{j'})^2 -  
(w_{j'}-\frac{w_{1,j} }{4})w_{1,j}  - \frac{2\e_{j-1}-w_{1,j}}{4  
w_{2,j}}  }
\\
- \frac{1}{g}+2w_{2,j} = 0.
\end{multline}
The resulting system of non-linear equations can then easily be solved  
using
Newton's method. Notice that every element of the Jacobian matrix has an
analytical expression that is easy to obtain and therefore is not  
explicitly
written here. Of course, for Newton's procedure to converge to the  
correct
solution at a given $g$, we need a good approximation to it. It is  
simple to
do so by slowly incrementing $g$ starting from $g=0$, where the GS is
known. One can then use a simple linear regression on $w_{1,j},w_{2,j}$  
to
obtain an educated guess to the ground state at $g' = g + \Delta g$.  
Despite
its simplicity this method, very similar to other ones in the literature
\cite{r-66,dr-01,rsd-03,s-07}, is sufficient for obtaining the ground  
state
solutions. For general states, for which the formation of and splitting  
apart
of CCPs can be highly non-trivial, a more refined algorithm (see Refs.
[\onlinecite{rnd-57,ded-06}] for example) is needed to find the  
solutions.

Fig. \ref{GSPLOT} shows three examples of the numerically computed  
ground
state solution of Richardson's equations. One can see that the generic
statements made about this solution in the preceding subsection are  
confirmed.
As $g$ gets sufficiently large and every rapidity has collapsed into a  
CCP (for
an even number of pairs), they arrange themselves into an arc in the  
complex
plane as shown more clearly in Fig. \ref{arc}.
This behavior was originally predicted
using the analogy between the set of equations and a two dimensional
electrostatic problem involving fixed and free charges
\cite{g-book,r-77,rsd-02}.

For a correct interpretation of the main features of the solutions to  
the
Richardson equations it is important to know the value of the  
superconducting
gap given by Eq. (\ref{Deltagen}) for the particular Hamiltonian we  
choose
(i.e. $\e_\a=\a$). It is easy to show that for large $N$
\be
\Delta=\frac{\Delta_{GC}}N = \frac1{2 \sinh 1/2g}\,,
\label{deltavsg}
\ee
an expression we will need to compare finite-size results with the  
grand-canonical ones.
Consequently Anderson's criterion \cite{a-59} for the presence
of superconductivity for large $N$ is
\be
\Delta \agt N \Rightarrow\, g\agt \frac1{2\ln 2N_r}\,,
\label{dellm}
\ee
showing the typical \cite {sild-01} logarithmic behavior of the small  
$g$
expansion.

\begin{figure}[t]
  \includegraphics[width=8.5cm]{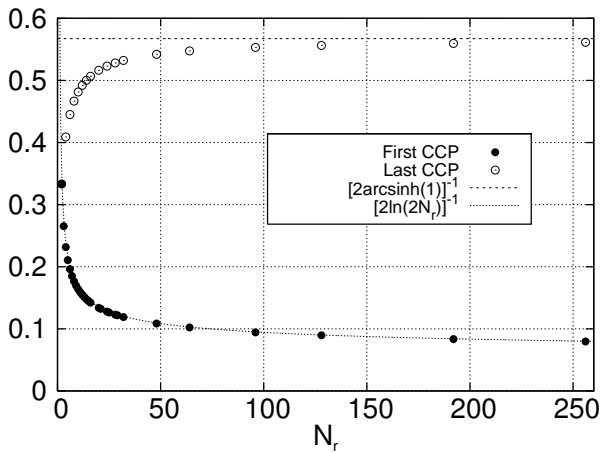}
\caption{Values of the coupling parameter at which the first and the  
last
complex conjugate pairs (CCP)  are formed}
\label{firstlast}
\end{figure}

Fig. \ref{firstlast} shows, as a function of $N_r$, the values of the  
coupling
constant $g^*_{N_r}(N_r)$ at  which the first two rapidities form a CCP.
We also plot, for even $N_r$, the values
of $g=g^*_1(N_r)$ at which the last couple of rapidities collapses into
a CCP. The latter is limited at large $N_r$ by \cite{rsd-02}
$g_0=(2\mathrm{arcsinh}1)^{-1} = 0.567296$ a  constant which is also
shown in the  figure.

These two numbers are particularly relevant to understand qualitatively  
the
different behaviors as function of $g$ and $N_r$.
In fact, when $g$ is larger than $g_1^*(N_r)$
all the particles are paired and the system has entered its asymptotic  
superconducting regime.
Oppositely when no pair has still been formed, i.e. for  
$g<g_{N_r}^*(N_r)$
superconductivity is absent. In fact, $g_{N_r}^*(N_r)$ coincides with
the critical value of the coupling given by Anderson's criterion
Eq. (\ref{dellm}) for large $N$.
The curve resulting from Eq. (\ref{dellm})
is plotted in Fig. \ref{firstlast} and the agreement with  
$g_{N_r}^*(N_r)$ is
excellent even for relatively small value of $N_r$.

Note also that $g_{N_r}^*(N_r)$ vanishes in the thermodynamic limit,  
which can
simply be interpreted as the Cooper instability.
A quantitative understanding of
these phenomena and of the crossover between small and large $g$ at  
fixed
finite $N_r$ requires an accurate study of the correlation functions,
which we present in the next section.

\subsection{Ground state energy}

\begin{figure}[t]
  \includegraphics[width=8.5cm]{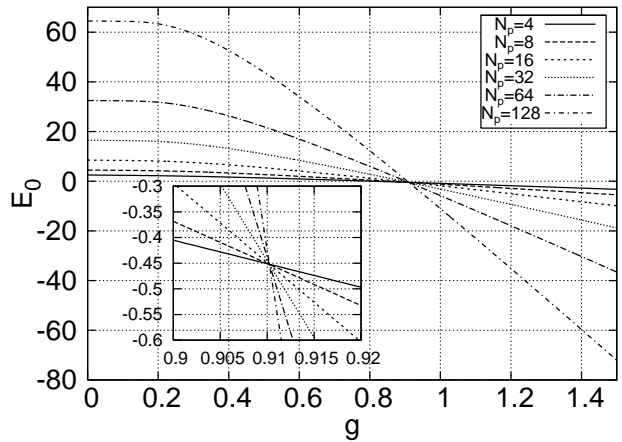}
\caption{Ground state energy per pair $E_0$ as a function of $g$.
Inset: Zoom close to the crossing point}
\label{e0}
\end{figure}

In Fig. \ref{e0} we plot the value of the ground state energy per
pair (in units of the inter-level spacing) at half-filling for a set of
different number of pairs as given by $N_p E_0=
  \sum_{j=1}^N\frac{\epsilon_j}{2}- \sum_j w_j$.

One interesting feature is the presence of a size invariant point at  
which
every curve cross (see the inset of Fig. \ref{e0} for a zoom close to
this point). Indeed at $g_{inv} \approx 0.910$ the ground state
energy seems to be independent of the number of pairs in the system
$E(g_{inv}) \approx -0.45$.
However, the presence of this ``fixed point'' does not carry any deep  
meaning
and can be easily understood in terms of the $1/N_p$ expansion developed
in Refs. [\onlinecite{r-77,yba-05}].
In fact, according to these references for large $N_p$ the ground-state
energy per particle can be written as
\be
E_0= N_p E_0^{(0)}+E_0^{(1)} +O(1/N_p)\,,
\label{E0exp}
\ee
with [$E_0^{(0)}$ is nothing but Eq. (\ref{E0rich})]
\bea
E_0^{(0)}&=&1-\frac12\coth\frac1{2g}\,,\\
E_0^{(1)}&=&\frac12(1-\phi(2g) \coth1/(2g)) \,,\\
\phi(2g)&=& \frac2\pi \int_0^\infty \frac{dx}{1+x^2}
\frac{\cosh\pi x/2}{\sqrt{\cosh^2(\pi x/2)+\sinh^2(1/(2g))}}\,,\nonumber
\eea
where we adapt the results to our normalization (i.e.
the quantities of Ref. [\onlinecite{yba-05}]
reads $D=2N_p$, $\lambda=2g$ and there
is a global shift of the energy levels).
The scale invariant point just corresponds to the value of $g$ for which
the order $N_p$ term $E_0^{(0)}$ vanishes, {\it i.e.}
$g_{inv} = (2 {\rm arccoth} 2)^{-1}= 0.910239\dots$.
The energy at this point, apart from $O(1/N_p)$ corrections,
is independent of $N_p$ and given by $E_0^{(1)}(g_{inv})=  
-0.45276\dots$.
Eq. (\ref{E0exp}) is thus practically a
perfect approximation of the actual value of the ground-state
energy for large enough $N_p$, say $N_p\geq 16$.

\begin{figure}[t]
\includegraphics[width=\columnwidth]{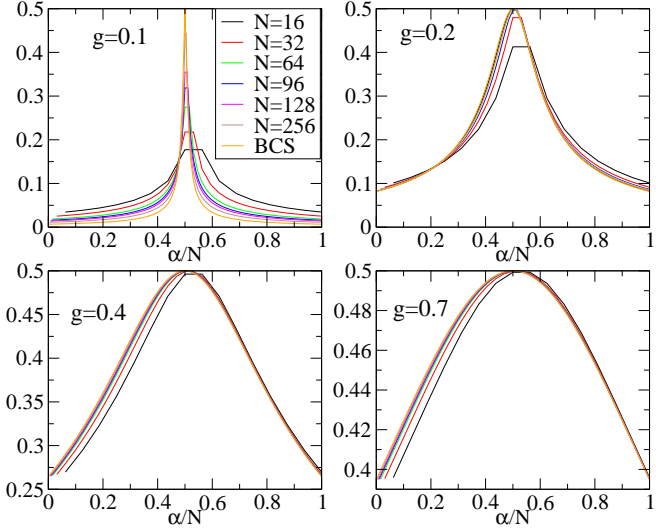}
\caption{(Color online) $u_\a v_\a$ as function of $\a$. Each plot is  
at fixed
   $g$ and for several $N_p=N/2$.}
\label{OPjvsN}
\end{figure}

\section{Calculation of the correlation functions}
\label{seccorr}

The formulas we obtained for the correlation functions are completely  
general
and are valid for any choice of the Hamiltonian parameters $\e_\a$ and  
$g$
(some care would have to be taken in the limit of coinciding energy  
levels, however).
To obtain a physical result we still have to perform the sum over the  
$N_r$
terms, introducing in the determinants for the form factors the solution
to the Richardson equations.
This cannot be done analytically, so we need to make a choice of the  
model to study.
As we already mentionned, we only consider the most-studied case in the
condensed matter literature, which consists of $N$ equidistant levels at
half-filling, i.e. $N=2N_r = 2N_p =N_f$.  We normalize the levels as
\be
\e_\a= \a\, \qquad {\rm with}\,\; \a=1\dots N\,,
\ee
i.e. we measure the energy scale in terms of the inter-level spacing and
we fix the Debye frequency (the largest energy level) to $N$.

\subsection{Correlations among the same level and ``canonical'' order
   parameter}

\begin{figure}[t]
\includegraphics[width=\columnwidth]{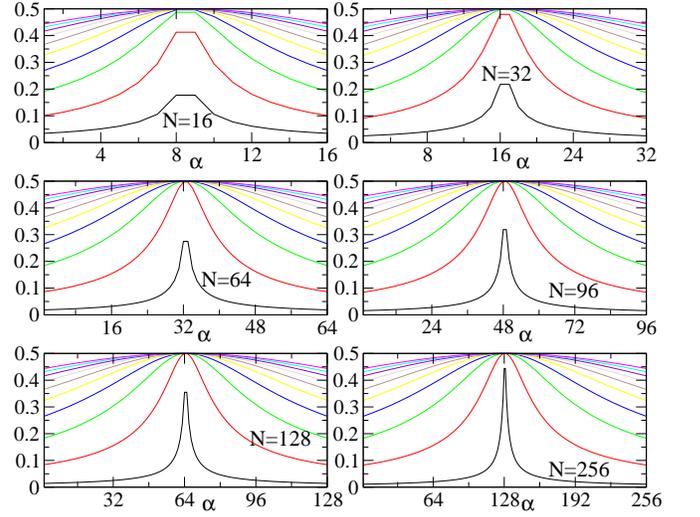} \vspace{2mm}
\caption{(Color online) $u_\a v_\a$ as function of $\a$. Each plot is  
at fixed
   $N_p$ but for several different $g$ going from $0.1$ (always the  
smallest) to
$1$ (always the largest) increasing by steps of $0.1$.}
\label{OPjvsg}
\end{figure}

Among the various correlation functions a central role is played by the  
ones on
the same level.
We consider the correlation
\be
u_\a v_\a=\sqrt{\langle S^-_\a S^+_\a\rangle \langle S^+_\a  
S^-_\a\rangle}=
\sqrt{1/4- \langle S^z_\a\rangle^2}\,,
\ee
that can be easily obtained by the previous representation of
$\langle S^z_\a\rangle$ and does not require the reduction formulas  
because it
is written in terms of a single form factor.
This correlation is important because it is one of the building blocks  
of the BCS
theory and because it allows to define a ``canonical'' BCS order  
parameter.
In fact, as already discussed, Eq. (\ref{DelGC}) defining the
grand-canonical gap, is always zero in the canonical ensemble.
Thus following Ref. [\onlinecite{dzgt-96}] we use as a
canonical order parameter
\be
\Psi=\sum_{\a=1}^N u_\a v_\a\,.
\ee
Note that $\Psi$ is just half of the concurrence (which is a
local entanglement measure, see as a review [\onlinecite{afov-07}])
which has been already calculated with the present method \cite{dlz-05}.

In the large $N$ limit all these correlators must reduce to the value  
in the
grand-canonical ensemble, which from Eq. (\ref{vjBCS}) specialized to  
$\e_a=\a$
is
\be
u_\a v_\a=\frac12 \frac{\Delta}{\sqrt{\Delta^2+(\a-N_p)^2/N^2}}\,,
\ee
where we fixed the chemical potential to $\mu=N/2=N_p$ and we recall  
that
$\Delta$ is given by Eq. (\ref{deltavsg}).
Consequently, in the same limit, the canonical order parameter is
\begin{multline}
\Psi_{N_p=\infty}= \lim_{N_p\to\infty} \sum_{\a=1}^{2N_p} u_\a v_\a=
\frac{N\Delta}4 \int_{-1}^1 \frac{dx}{\sqrt{\Delta^2+(x/2)^2}}=\\
\frac{N\Delta}2 \log \frac{\sqrt{1+4\Delta^2}+1}{\sqrt{1+4\Delta^2}-1}=
\frac{N\Delta}{2 g}=\frac{N}{4 g \sinh1/2g}\,.
\label{BCSOP}
\end{multline}
It is evident that $\Psi$ vanishes when the gap $\Delta$ is zero,  
confirming
that in the thermodynamic limit it is a good order parameter.

Our results for $u_\a v_\a$ are reported in Figs. \ref{OPjvsN}
and \ref{OPjvsg}. In the former each plot consists of the various curves
at fixed $g$ (=0.1, 0.2, 0.4, 0.7) with varying $N_p$.
The latter instead shows the $g$ dependence at fixed $N_p$.
In Fig. \ref{OPjvsN} also the BCS results for any $g$
are reported for comparison.
It is evident that for all $g$ the results tend to converge to the BCS  
ones,
as they must.
However this convergence is slower as $g$ is smaller, for example for  
$g=0.1$
the maximum at $N=256$ is only 90\% of the asymptotic result and
conversely at $g=0.7$ the $N_p=16$ result is already 99.8\%.
These finite $N_p$ correlations are symmetric with respect to $(N+1)/2$
by construction. However we point out that this will not be true for
different level correlations, while in the grand-canonical ensemble
they are both symmetric.

\begin{figure}[t]
\vspace{5mm}
\includegraphics[width=0.9\columnwidth]{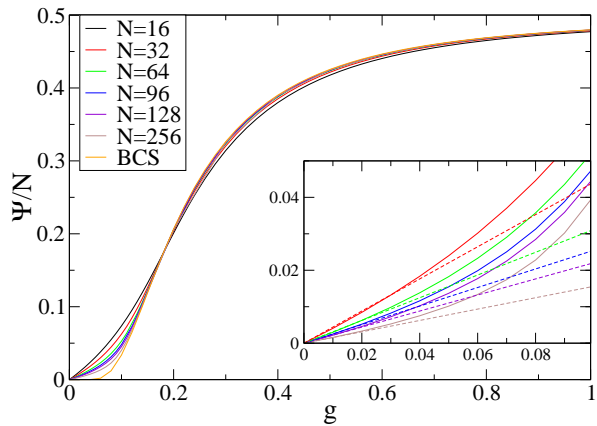}
\caption{(Color online) Canonical order parameter $\Psi$ as a function  
of
$g$ for different numbers of pairs $N_p=N/2$.}
\label{figPsi}
\end{figure}

In Fig. \ref{figPsi} we report the order parameter $\Psi/N$ as a  
function of
$g$ for several values of $N_p$, and compare it with the BCS result.
This figure is exactly the same as the one for the concurrence obtained
by Dunning et al. \cite{dlz-05}, with the important difference that they
considered only $N_p\leq34$ while we pushed the calculation up to  
$N_p=128$.
We could have calculated these correlations for larger $N_p$,
but the ones considered are already enough to describe
the crossover from the mesoscopic to the macroscopic regime.
In fact, Fig. \ref{figPsi} shows that for $N_p=128$ $\Psi$ is almost
indistinguishable from the BCS one Eq. (\ref{BCSOP}), except for very  
small
$g$ that are characterized by the scaling (\ref{dellm}).

\begin{figure}[t]
\vspace{2mm}
\includegraphics[width=0.95\columnwidth]{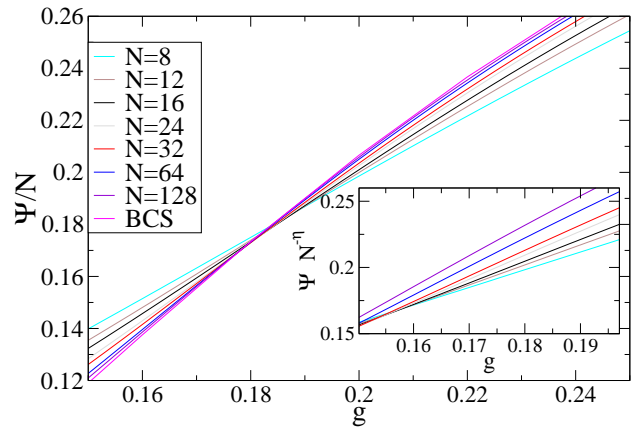}
\caption{(Color online) Zoom of $\Psi$ in the region $0.15<g<0.25$ for
several $N_p=N/2$. Inset:
Scaling ansatz of reference [\onlinecite{mff-98,ao-02}] and its failure
for large $N_p$.}
\label{figPsizoom}
\end{figure}

 >From the figure, it is evident
that for $g<g^*\sim0.18$ the BCS limit is reached from above, whereas  
for
$g>g^*$ it is approached from below. Exactly at $g=g^*$ all the curves  
seem to
cross in the same point. This is slightly different from what was  
observed
before for a small number of particles \cite{mff-98,ao-02},
where to get a similar crossing the
order parameter was multiplied by $N^{-\eta}$ with $\eta\simeq 0.94$.
To clarify this point we zoom in on the crossing point in
Fig. \ref{figPsizoom}. It is evident that for $N_p\geq 8$ all the curves
approximately cross in $g^*$, but this is not the case for smaller  
sizes.
It is then direct to interpret $g^*$ as the value of $g$ where the  
leading
finite-size correction of order $1/N$ vanishes (in fact these are  
clearly
negative for large $g$ and positive for very small ones).
The differences for smaller size are due to higher
order corrections $\sim 1/N^2$.
This fixed point is thus completely analogous to results discussed in  
the
previous section for the ground-state energy.
A very interesting problem would be to calculate $g^*$ directly from the
finite-size form in an analytical manner using the $1/N_p$ expansion  
previously
discussed \cite{r-77,yba-05}.

The finite-size scaling $\Psi\sim N^{\eta}$ found in Refs.
[\onlinecite{mff-98,ao-02}]
can clearly not be true for large sizes, since $\Psi$ is an extensive
quantity.
To check for which sizes it stops working,
in the inset of Fig. \ref{figPsizoom} we
plot $\Psi N^{\eta}$. All the systems with sizes $N_p \leq 16$ cross  
indeed at
the value of Refs. [\onlinecite{mff-98,ao-02}]
$g_{\rm cr}\sim 0.157$, but larger systems clearly deviate from this  
fixed
point. We can then safely conclude
that this scaling ansatz is effective only for $N_p \leq 16$.
In Ref. [\onlinecite{ao-02}] a second crossing point has been also  
found for a
larger value of the pairing constant. According to our analysis also  
this
fixed point is present only for relatively small number of pairs.

For large $g$ the BCS result is the leading term for large $N$ and  
easily
gives $\Psi/{N}=1/2-1/({48 g^2})+O(g^{-3},N^{-1})$ whereas for small
coupling we have \cite{sild-01}
\be
\frac\Psi{N}=
g\frac{\ln (3+\sqrt8))}{\sqrt{N_p}}+O(1/\ln N_p)\,,\quad {\rm for}\;  
g\ll1\,.
\ee
Note that for large $g$ we have an $N_p$ independent result while for  
small $g$
there is a square-root singularity in $N_p$. The latter is again a  
manifestation of the
non-perturbative nature of superconductivity.
Both these analytical results are perfectly reproduced by our numerics.

\subsection{Static correlation functions among different levels}

\begin{figure}[t]
\includegraphics[width=\columnwidth]{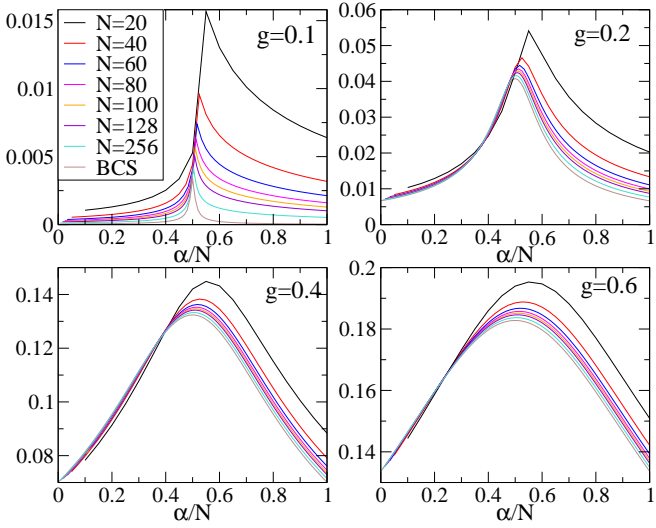} \vspace{2mm}
\caption{(Color online) Off-diagonal correlations
$\langle S^-_1 S^+_\a\rangle$ as a function of $\a/N$.}\label{S+1S-n}
\end{figure}

\begin{figure}[t]
\includegraphics[width=\columnwidth]{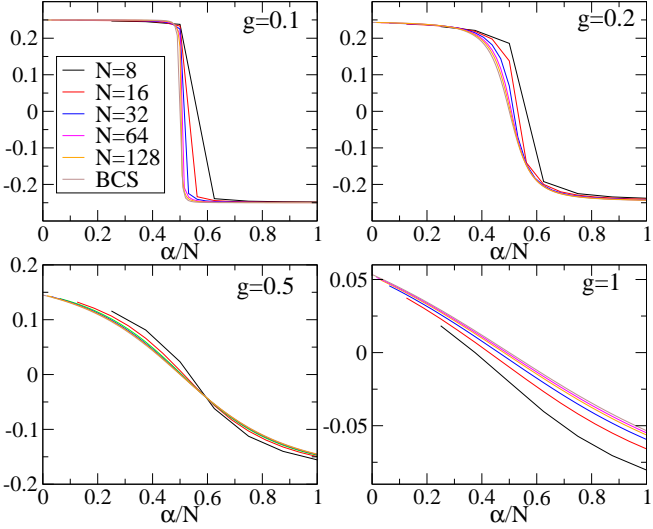} \vspace{2mm}
\caption{(Color online) $\langle S^z_1 S^z_\a\rangle$ as function of  
$\a/N$.}
\label{Sz1Szn}
\end{figure}

Despite several interesting features of the correlation functions among  
the
same levels that we have just discussed, these are qualitatively very  
similar
to the grand-canonical ones.  On the other hand,
correlation functions between different levels (known as off-diagonal
ones) are a strong signature of the canonical BCS-like pairing  
correlations
and should be relevant for the interpretation of tunneling experiments.
In fact, within the grand-canonical ensemble (and so for $N=\infty$)
these four-point correlation functions factorize to the product of
two point ones (i.e. in this approximation the Cooper pairs are free).
Oppositely, in the canonical ensemble they are non-trivial functions of
both the pairs as a consequence of quantum fluctuations.
Following Ref. [\onlinecite{ao-02}], we concentrate here on the two  
correlation
functions
\be
\langle S^-_1 S^+_\a \rangle\,, \quad {\rm and}\quad
\langle S^z_1 S^z_\a \rangle\,.
\ee
Our results for different values of $g$ and $N$ are reported in
Fig. \ref{S+1S-n} and \ref{Sz1Szn} respectively.

\begin{figure}[t]
\includegraphics[width=\columnwidth]{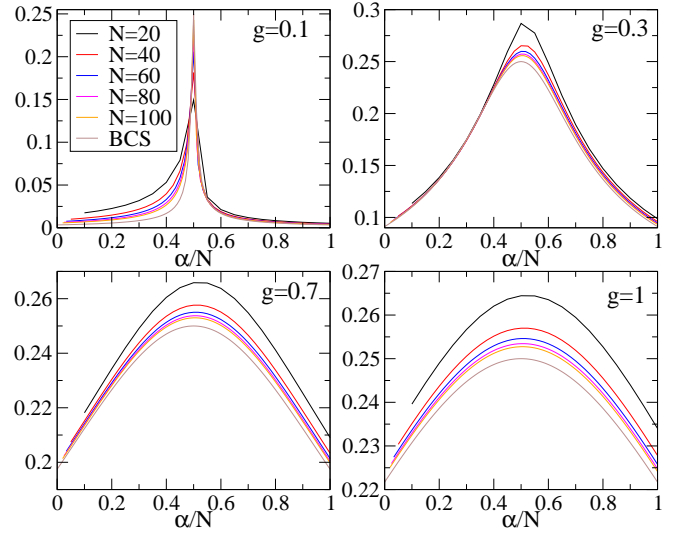} \vspace{2mm}
\caption{(Color online) $\langle S^-_{N_p+1} S^+_\a\rangle$
as function of $\a/N$.}\label{S+N2S-n}
\end{figure}

For $N\to\infty$, as a consequence of factorization, we have
\begin{multline}
\langle S^-_\a S^+_\b \rangle= u_\a v_\a u_\b v_\b= \\
\frac14 \frac{\Delta}{\sqrt{\Delta^2+(\a-N_p)^2/N^2}}
\frac{\Delta}{\sqrt{\Delta^2+(\b-N_p)^2/N^2}}\,.
\end{multline}
In particular at fixed $\beta$ these correlations are symmetric with  
respect
to $\a=N_p$. Including some trivial finite size effect, in the  
grand-canonical
ensemble one would expect a
symmetry at $(N+1)/2$ as for the on-level correlations.
However, as evident from the figures this is not the case in the  
canonical
description. Furthermore, the smaller $g$ is the more
asymmetrical are the correlations. Such asymmetries are very pronounced  
for
all the $S^-S^+$ correlators, as for example showed in Fig.  
\ref{S+N2S-n}
where we report as the other extreme (compared to
$\langle S^-_1 S^+_\a \rangle$),
the correlator $\langle S^-_{N_p+1} S^+_\a \rangle$.
Thus the asymmetries can be used to
understand the degree of ``canonicality'' of a system.
Note in particular the very different scales in Figs. \ref{S+1S-n} and
\ref{S+N2S-n}: off-diagonal correlation functions are much more  
important when
one of the levels is close to the Fermi point, a fact that is not  
surprising
being true also in the grand-canonical ensemble.

A last property that is not apparent from the plots but that is true
(even if not evident from the determinant representation) is that
\be
\langle S^-_\a S^+_\b \rangle=\langle S^-_\b S^+_\a \rangle\,,
\ee
that we checked for all the values we calculated.

The correlation function $S^z_\a S^z_\b$ in the grand-canonical ensemble
also factorizes into the product  of two point ones:
\be
\langle S^z_\a S^z_\b \rangle=
\frac{(\a-N_p)/2N }{\sqrt{\Delta^2+(\a-N_p)^2/N^2}}
\frac{(\b-N_p)/2N}{\sqrt{\Delta^2+(\b-N_p)^2/N^2}}\,,
\ee
and it is an odd function at $\a=M$ (or $\b$).
Again the finite $N$ results do not have this symmetry, that is  
recovered only in
the thermodynamic limit.
The smoothing of the step-like structure increasing $g$ is a well-known  
effect
also in the grand-canonical ensemble.

\subsection{Off-diagonal order parameter}

\begin{figure}[t]
  \includegraphics[width=8.5cm]{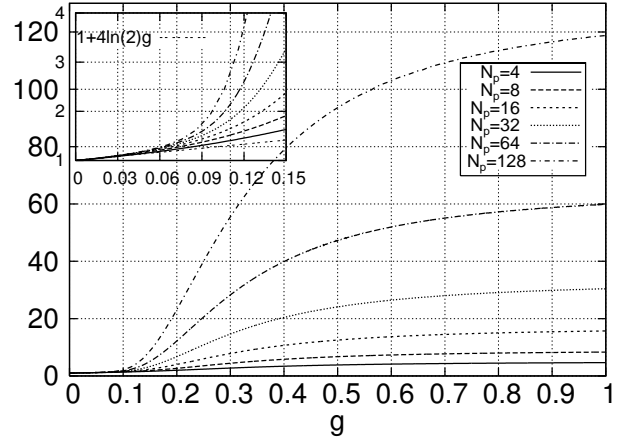}
\caption{Summed correlation function
$\frac{1}{N_p} \sum^N_{\a,\b=1}  \left< S^-_\a S^+_\b\right>$ as a  
function of
$g$ for various number of pairs.
Inset: small $g$ behavior compared with the analytic expression.}
\label{de0}
\end{figure}

Another fundamental quantity is the so called off-diagonal long-range  
order
parameter\cite{po-56} defined by
\be
\Psi_{OD} \equiv \frac{1}{N_p} \sum^N_{\a,\b=1} \left< S^+_\a  
S^-_\b\right>\,,
\ee
that as a difference with $\Psi$ takes into account the effect of  
non-diagonal
correlations.
$\Psi_{OD}$ is clearly accessible from the direct computation of the
off-diagonal correlation functions (as already proposed \cite{zlmg-02}),
but this is not needed. In fact, it can be obtained without the use of  
the
determinant representation, using the Hellmann-Feynman theorem (an  
alternative
method of calculation has been also proposed \cite{od-05}).
The derivative of the ground-state energy with respect to the
coupling strength, allows us to directly compute the double sum over
all levels of the static $S^+S^-$ correlation  function, i.e.
\be
\Psi_{OD}=
\frac{1}{N_p} \sum^N_{\a,\b=1} \left< S^+_\a S^-_\b\right>=
- \frac{1}{N_p} \frac{\partial E_0(g)}{\partial g}\,.
\label{FVth}
\ee
Fig. \ref{de0} shows this summed correlation for different pair
numbers. At $g=0$, the only contributing terms are the one for which
$\b=\alpha$ since no correlations between pairs in different  levels  
exist and
we trivially have  $\sum^N_{\a,\b=1} \left< S^+_\a S^-_\b\right>  =  
N_p$.
As the interaction is turned on inter-level correlations build up  
rapidly until they
saturate for maximally correlated wavefunctions.  For every $N_p$, this
large $g$ limit is clearly given  by $\Psi_{OD}=N_p+1$.

The small $g$ behavior can be obtained analytically from the known  
result for
the energy \cite{sild-01}
\begin{multline}
{1-E_0}= g +2 g^2 \ln 2 +O(g^3,(\ln L)^{-1})\\
\Rightarrow \Psi_{OD}= 1+g 4\ln2 +O(g^2)\,.
\end{multline}
This curve is shown in the inset of Fig. \ref{de0} and perfectly agrees  
with
the numerical results.
Again the deviations from this behavior start to occur at a value of  
$g$ given by the
usual logarithmic scaling of Eq. (\ref{dellm}).

In the thermodynamic limit, as a consequence of the factorization,  
$\Psi_{OD}$
is trivially related to $\Psi$ as
\be
\Psi_{OD}^{N_p=\infty}=\frac{\Psi^2_{N_p=\infty}}{N_p}\,,
\ee
signaling that one is extensive if and only if the other one is.  
However, as
evident from the figure, for fixed finite $N_p$ this is not true and  
the two
quantities are independent.
Furthermore for small $g$, in the regime that is not accessible by the
BCS ansatz, they are both linear in $g$ and cannot be in a quadratic
relation as for large $N_p$.
Actually, it has been proven \cite{ttc-00} that $\Psi$ and $\Psi_{OD}$
satisfy the following relations for any value of $g$ and $N_p$
\be
\frac{1}{N_p} \Psi(\Psi-1)\leq \Psi_{OD}\leq1+\frac{N}{N_p} \Psi\,,
\ee
that for $N_p\to\infty$ are trivial bounds, but not for finite $N_p$.
We checked that our calculations satisfy these bounds.

The direct knowledge of $\Psi_{OD}$ allows for a
last, very important consistency check.
In fact the value found from Hellmann-Feynman theorem must equal, at  
half-filling, the sum
$\frac{1}{N_p} \sum^N_{\a,\b=1}  \left< S^-_\a S^+_\b\right>$
calculated from the determinant representation.
We checked for all $N_p\leq 64$ that this is indeed the case.

\section{Conclusions}
\label{concl}

We have studied the static correlation functions of the reduced BCS  
model in the
canonical ensemble. From the theoretical point of view we simplified the
results of Ref. [\onlinecite{zlmg-02}]
giving the correlations as sums over only $N_r$
determinants of $N_r\times N_r$ matrices. This allowed us to calculate  
the
correlation functions for very large numbers of particles, describing
the crossover from mesoscopic to the thermodynamic limit, and going
beyond previous exact or approximate studies.
In particular with such accurate calculations we were able to discuss
critically some conjectured scaling forms for the canonical order  
parameter.
For example we rule out the idea of any phase transition as a function  
of the
(positive) pairing strength and number of particles, in agreement with
other analyses based on thermodynamical quantities  
\cite{ds-99,sdddb-00,dr-01}.
We also calculate the off-diagonal long-range order parameter by using  
the
Hellmann-Feynman theorem.

A first interesting step to go beyond what has been done here would be  
to find a single
determinant representation for the correlation functions. We made  
several
attempts in this direction, but so far unsuccessfully.

We only analyzed the case of $N$ non-degenerate equidistant energy  
levels at
half-filling. This is the most interesting model from the
condensed matter point of view. However for the description of pairing  
in
nuclei other choices of the parameters $\e_\a$ are more
natural \cite{dh-03,bdh-04,zv-05,sv-07}.  These could be treated by
a simple adaptation of our results.

Furthermore the method presented here allows in principle to obtain  
dynamical
correlation functions as sums of form factors over the excited states.
This is a more numerically demanding problem, but can be tackled in the  
same way
as other Bethe Ansatz solvable models \cite{sc,1dbg}.
Also, by summing over the excited states, one can access the finite
temperature thermodynamics and this can help in understanding some open
questions \cite{lfh-00,sv-07} for ultra-small metallic grains.

The Hamiltonian (\ref{BCSH}) is the simplest one with pairing terms.  
More
general models with several couplings have been proposed to explain
complicated pairing in condensed matter
\cite{gcm} and nuclear physics \cite{gnp}.
Some have been also shown to be integrable \cite{ig,lzmg-03}.
As a consequence it would be extremely interesting to tackle these  
models
with methods similar to those presented here.

\section*{Acknowledgments}

We thank Luigi Amico for extremely fruitful discussions and  
correspondence.
All the authors are thankful for support from the Stichting voor
Fundamenteel Onderzoek der Materie (FOM) in the Netherlands.
This work has been done mainly when PC was a guest of
the Institute for Theoretical Physics of the Universiteit van
Amsterdam. This stay was supported by the ESF Exchange Grant 1311 of
the INSTANS activity.


\begin{thebibliography}{999}

\bibitem{Rajagopalh0011333} K. Rajagopal and F. Wilczek, hep-ph/0011333.

\bibitem{AlfordARNPS51} M. Alford, Ann. Rev. Nucl. Part. Sci. 51, 131  
(2001).

\bibitem{bcs-57}
J. Bardeen, L. N. Cooper, and J. R. Schrieffer,
Phys. Rev. {\bf 106}, 162 (1957);  {\it ibid.} {\bf 108}, 1175 (1957).

\bibitem{rs-62}
R. W. Richardson, Phys. Lett. {\bf 3}, 277 (1963); {\bf 5}, 82 (1963);
R. W. Richardson and N. Sherman, Nucl. Phys. {\bf 52}, 221 (1964);
{\bf 52}, 253 (1964).

\bibitem{exp}
D. C. Ralph, C. T. Black, and M. Tinkham,
Phys. Rev. Lett. {\bf 74}, 3241 (1995);
and ibid. {\bf 76}, 688 (1996);
and ibid. {\bf 78}, 4087 (1997).

\bibitem{dr-01}
J. von Delft and D. C. Ralph, Phys. Rep. {\bf 345}, 61 (2001).

\bibitem{a-59}
P. W. Anderson, J. Phys. Chem. Solids {\bf 11}, 28 (1959).

\bibitem{dps-04} J. Dukelsky, S. Pittel, and G. Sierra,
Rev. Mod. Phys. {\bf 76},  643 (2004)

\bibitem{dh-03} D. J. Dean and M. Hjorth-Jensen,
Rev. Mod. Phys. {\bf 75}, 607 (2003)

\bibitem{r-65}
R. W. Richardson, J. Math. Phys. {\bf 6}, 1034 (1965).

\bibitem{ao-02}
L. Amico and A. Osterloh, Phys. Rev. Lett. {\bf 88}, 127003 (2002).

\bibitem{sk}
E. K. Sklyanin, Lett. Math. Phys. {\bf 47}, 275 (1999).

\bibitem{zlmg-02}
H.-Q. Zhou, J. Links, R.H. McKenzie, and M.D. Gould,
Phys. Rev. B {\bf 65}, 060502(R) (2002).

\bibitem{lzmg-03}
J. Links, H.-Q. Zhou, R.H. McKenzie, and M.D. Gould,
J. Phys. A {\bf 36}, R63 (2003).

\bibitem{s-89}
N. A. Slavnov, Teor. Mat. Fiz. {\bf 79}, 232 (1989).

\bibitem{dlz-05}
C. Dunning, J. Links, and  H.-Q. Zhou,
Phys. Rev. Lett. {\bf 94}, 227002 (2005).

\bibitem{mff-98}
A. Mastellone, G. Falci, and R. Fazio, Phys. Rev. Lett. {\bf 80}, 4542  
(1998).

\bibitem{g-book}
M. Gaudin,
{\it Mod\`eles Exactement R\'esolus} (Les \'Editions de Physique,
Les Ulis, France, 1995).

\bibitem{crs-97} M. C. Cambiaggio, A. M. F. Rivas, and M. Saraceno,
Nucl. Phys. A {\bf 624}, 157 (1997).

\bibitem{aff-01} L. Amico, G. Falci, and R. Fazio,
J. Phys. A {\bf 34} 6425, (2001).

\bibitem{dp-02}
J. von Delft and R. Poghossian, Phys. Rev. B {\bf 66}, 134502 (2002)

\bibitem{r-77} R.W. Richardson, J. Math. Phys. {\bf 18}, 1802 (1977).

\bibitem{r-66}
R. W. Richardson, Phys. Rev. {\bf 141}, 949 (1966).

\bibitem{sild-01}
M. Schechter, Y. Imry, Y. Levinson, and J. von Delft,
Phys. Rev. B {\bf 63}, 214518 (2001).

\bibitem{yba-03}
E. A. Yuzbashyan, A. A. Baytin, and B. L. Altshuler,
Phys. Rev. B {\bf 68}, 214509 (2003).

\bibitem{sg-06}
I. Snyman and H. B. Geyer, Phys. Rev. B {\bf 73}, 144516 (2006).

\bibitem{rsd-03}
J. M. Roman, G. Sierra, and J. Dukelsky, Phys. Rev. B {\bf 67} 064510  
(2003).

\bibitem{s-07}
M. Sambataro, Phys. Rev. C {\bf 75}, 054314 (2007).

\bibitem{rnd-57}
S. Rombouts, D. Van Neck and J. Dukelsky, Phys. Rev. C {\bf 69}, 061303  
(2004).

\bibitem{ded-06}
F. Dominguez, C. Esebbag, and J. Dukelsky, J. Phys. A {\bf 39}, 11349  
(2006).

\bibitem{rsd-02}
J.M. Roman, G. Sierra, and J. Dukelsky, Nucl.Phys. B {\bf 634}, 483  
(2002).



\bibitem{yba-05}
E. A. Yuzbashyan, A. A. Baytin, and B. L. Altshuler,
Phys. Rev. B {\bf 71}, 094505 (2005)







\bibitem{dzgt-96}
J. von Delft, A. D. Zaikin, D. S. Golubev, and W. Tichy,
Phys. Rev. Lett. {\bf 77}, 3189 (1996)

\bibitem{afov-07}
L. Amico, R. Fazio, A. Osterloh, and V. Vedral,
Rev. Mod. Phys., to appear [quant-ph/0703044].

\bibitem{po-56}
O. Penrose and L. Onsager, Phys. Rev. {\bf 104}, 576 (1956);
C. N. Yang, Rev. Mod. Phys. {\bf 34}, 691 (1962).

\bibitem{od-05}
G. Ortiz and J. Dukelsky,
Phys. Rev. A {\bf 72}, 043611 (2005).

\bibitem{ttc-00}
G.-S. Tian, L.-H. Tang, and Q.-H. Chen, Europhys. Lett. {\bf 50}, 361  
(2000);
Phys. Rev. B {\bf 63}, 054511 (2001).

\bibitem{ds-99}
J. Dukelsky and G. Sierra, Phys. Rev. Lett. {\bf 83}, 172 (1999)

\bibitem{sdddb-00}
G. Sierra, J. Dukelsky, G. G. Dussel, J. von Delft, F. Braun,
Phys. Rev. B {\bf 61}, 11890 (2000).

\bibitem{bdh-04}
A. Belic, D.J. Dean, and M. Hjorth-Jensen,
Nucl. Phys. A {\bf 731}, 381 (2004).

\bibitem{zv-05}
V. Zelevinsky and A. Volya, Nucl. Phys. A {\bf 752}, 325 (2005).

\bibitem{sc}
J.-S. Caux and J. M. Maillet, Phys. Rev. Lett. {\bf 95}, 077201 (2005)
J.-S. Caux, R. Hagemans, and J. M. Maillet, J. Stat. Mech. P09003  
(2005).

\bibitem{1dbg}
J.-S. Caux and P. Calabrese, Phys. Rev. A {\bf 74}, 031605R (2006);
J.-S. Caux, P. Calabrese, and N. A. Slavnov, J. Stat. Mech. P01008  
(2007).

\bibitem{sv-07}
T. Sumaryada and A. Voyla, Phys. Rev. C {\bf 76}, 024319 (2007).

\bibitem{lfh-00}
A. Di Lorenzo, Rosario Fazio, F.W.J. Hekking, G. Falci, A. Mastellone,  
and
G. Giaquinta, Phys. Rev. Lett. {\bf 84}, 550 (2000).

\bibitem{gcm}
I. L. Kurland, I. L. Aleiner, and B. L. Altshuler,
Phys. Rev. B {\bf 62} 14886 (2000);
J. Dukelsky, G. G. Dussel, C. Esebbag, and S. Pittel,
Phys. Rev. Lett. {\bf 93}, 050403 (2004);
J. Dukelsky, G. Ortiz, S.M.A. Rombouts, and K. Van Houcke
Phys. Rev. Lett. {\bf 96}, 180404 (2006);
A. M. Garc\'ia-Garc\'ia, J. D. Urbina, E. A. Yuzbashyan, K. Richter and  
B. L. Altshuler,
arXiv:0710.2286.


\bibitem{gnp}
J. Dukelsky, V. G. Gueorguiev, P. Van Isacker, S. Dimitrova, B. Errea,  
and
S. Lerma H, Phys. Rev. Lett. {\bf 96} 072503 (2006);
S. Lerma H., B. Errea, J. Dukelsky, and W. Satula,
Phys. Rev. Lett. {\bf 99}, 032501 (2007).


\bibitem{ig}
L. Amico, A. Di Lorenzo, and A. Osterloh,
Phys. Rev. Lett. {\bf 86}, 5759 (2001);
J. Links, H.-Q. Zhou, R.H. McKenzie, and M.D. Gould
Int. J. Mod. Phys. B {\bf 16}, 3429 (2002).

\end{thebibliography}
\end{document}